# A New computation reduction based nonlinear Kalman filter

M.Behvandi
Student
Electrical department
Amirkabir university of technology
Tehran, Iran
Email: behvandi@aut.ac.ir

A.A. Suratgar
Associated Professor
Department
Amirkabir university of technology
Tehran, Iran
Email: a-suratgar@aut.ac.ir

M.A.Khosravi
Assisted Professor
Electrical department
Amirkabir university of technology
Tehran, Iran
Email: m.a.khosravi@aut.ac.ir

*Abstract*— **This article introduces a new algorithm for nonlinear state estimation based on deterministic sigma point and EKF linearized framework for priori mean and covariance respectively. This method reduces the computation cost of UKF about 50% and has better accuracy compared to EKF due to propagating mean and Covariance of state to 3rd order Taylor series. Several types of Kalman filter have been presented before to reduce the computation cost of UKF, however, this new KF is a better choice because of its simplicity, numerical stability and accuracy for real time implementation. Examples verify the effectiveness of the proposed method.**

*Keywords-component; Nonlinear state estimation, UKF, EKF, Computation cost, Unscented transformation*

## I. INTRODUCTION

Kalman filter is widely used in state estimation, maneuver tracking, control, slam, image processing and etc.... To deal with nonlinear system EKF undoubtedly is the most widely used nonlinear state estimation technique that has been applied in the past few decades. EKF use linearized method to propagate mean and covariance error, and in some cases goes to divergence because of linearized error. EKF use first order Taylor series to estimate mean, and it is accurate to 3rd order for covariance estimation. In [1] Julier and Ulham used sigma points to transform mean, and covariance through nonlinear function to obtain better results in mean and covariance and called the method unscented transformation (UT). They showed that the unscented Kalman filter is more accurate comparing to EKF and provides a better approximation of mean and covariance of the state. UKF approximates mean and covariance both to 3rd order Taylor series term. UKF needs more computation time and is numerically unstable with respect to round off error [17], [18] that make EKF first choice for real time applications.
In many research's, special attention has been paid to the KF and its derivative for designing effective filter implementations that increase robustness of the filters against round off error. The most popular and useful techniques are found in the class of square root (SR) or UDU$^T$ factorization-based (UD-based) methods. To solve the computation time several methods have been used, Julier presented reduced order sigma point UKFs called them simplex unscented Kalman filter and spherical simplex unscented Kalman filter [5]- [6]. The reduced sigma point Kalman filters reduce the computation of UKF less than 50% in maximum and have an accuracy to second order Taylor series [5]. simplex UKF uses $n+1$ sigma point but has a stability problem due to its sigma point radius related to $2^{n/2}$ [5],[6]. To concern stability of filter spherical simplex UKF proposed in [6] n+1 sigma points lie on hypersphere radius which is proportional to $\sqrt{n}$ . In [2] Biswas et all proposed new method called Single Point UKF (SPUKF) that reduced computation cost without reducing sigma points. The idea was to propagate posterior mean and calculate other sigma points using first order Taylor series approximation. The approximate mean and covariance are accurate to first order Taylor series of original UT[1] which makes it a little better than EKF in the mean estimation and covariance. They assumed that it is necessary to use 4 orders Rung Kutta method to solve differential equation with step size h>20 in each propagation. So their approach has a good performance to reduce computation under two assumptions:

1- The System is a highly nonlinear continuous system, that needs Rung-Kutta numerical integration.

2- To obtain better accuracy of rung Kutta method, $\delta t / h$ ( $h \in N$ ) step size is required to propagate state vector from time $t$ to $t + \delta t$ .

These assumptions mean that in each iteration nonlinear function of the system should be evaluated by (2n+1) h*4 times. This has an extra computation cost in real time application, though it is not necessary in most actual cases. All these methods reduce the accuracy of UKF to reduce computation cost so that they are useful when computation cost is the main problem.

---

[1] unscented transformation

In [9] Chang it was suggested the Marginal Unscented Transformation (MUT) to reduce the number of sigma points, where it can be applied to a special type of nonlinear function containing linear substructures. It was suggested that, if $n_a$ state elements out of the $n$ state elements are mapped non-linearly then the number of sigma points can be reduced to $2n_a + 1$. MUT can be applied only to the systems which contain linear substructures, so the computation complexity of this method dependent to the linear term of state space model.

In this paper, we want to propose a new Kalman filter which estimates mean and covariance of state to 3rd order Taylor series of true mean and covariance without reducing sigma point or reducing the performance of UKF. This method reduces computation cost of filter about 50%. The new method uses UT to obtain mean and EKF framework to estimate matrix covariance. As regards that EKF has the same accuracy in covariance estimation with less computation cost, it is expected that New UKF performs like the UKF with less computing time.

The rest of the paper is organized as follows: section II compare UT and linearized approximation of mean and covariance with true nonlinear transformation mean and covariance. In section III, the new UKF algorithm is described. In section IV computation analyses of New UKF, SPUKF and original UKF are shown. In section V New UKF, UKF, SSUKF, EKF are applied to (example 1), and (example 2) as benchmarks.

## II. BACKGROUND

Approximation approaches of mean and covariance of states are the main difference between EKF and UKF that leads to a better performance of UKF. In this section, the true mean and covariance of nonlinear function will be discussed.

In the next section, we lay out the basic framework of the analysis, which is the Taylor Series. In the subsequent sections, we examine the prediction of the mean and the covariance.

Let x be as a mean "$\bar{x}$" plus a zero-mean disturbance $\Delta x$ with covariance P, and the Taylor series for the non-linear transformation g[x], about $\bar{x}$ is:

$$g[\bar{x} + \Delta x] = g[\bar{x}] + D_{\Delta x}g + \frac{D_{\Delta x}^2 g}{2!} + \frac{D_{\Delta x}^3 g}{3!} + \frac{D_{\Delta x}^4 g}{4!} + ... \quad (1)$$

State prediction could be calculated regarding the expectation of Equation (1).

$$\bar{y} = E[g[\bar{x} + \Delta x]]$$
$$= g[\bar{x}] + E\left[D_{\Delta x}g + \frac{D_{\Delta x}^2 g}{2!} + \frac{D_{\Delta x}^3 g}{3!} + \frac{D_{\Delta x}^4 g}{4!} + ...\right] \quad (2)$$

For the true mean, denoted by the subscript T, $\Delta x$ is a zero-mean, Gaussian process with covariance P. By symmetry, all the odd ordered moments in this distribution are zero. Therefore, the expected value of all the odd terms in this series are zero and

$$\bar{y}_T = g[\bar{x}] + E\left[\frac{D_{\Delta x}^2 g}{2!} + \frac{D_{\Delta x}^4 g}{4!} + ...\right] \quad (3)$$

Since $E[\Delta x \Delta x^T] = P_{xx}$ the second order even terms can be written as

$$E\left[\frac{D_{\Delta x}^2 g}{2!}\right] = \left(\frac{\nabla^T P_{xx} \nabla}{2!}\right)g \quad (4)$$

The equation (2) can be written as

$$\bar{y}_T = g[\bar{x}] + \left(\frac{\nabla^T P_{xx} \nabla}{2!}\right)g + E\left[\frac{D_{\Delta x}^4 g}{4!} + ...\right] \quad (5)$$

EKF truncates this series at the first order and predicts the conditional mean as:

$$\bar{y}_{EKF} = g[\bar{x}] \quad (6)$$

This estimate is independent of the covariance and higher moments of the distribution of $\bar{x}$. However, comparing this with Equation (5) reveals that it is accurate only if the expected value of the second and higher order terms in the series are zero. This is always true for a linear system since the second and higher derivatives of the transformation are zero. However, for a general nonlinear system, these terms are non-zero and this condition does not hold. Therefore, errors are introduced in the second order.

The unscented Kalman filter predicts the mean from the projected set of points. Consider the Taylor series for the transition of each point $\chi_i$. This can be expressed as the Taylor series about $\bar{x}$.

$$\bar{y} = g[\bar{x}] + \frac{1}{2(n+k)}\sum_{i=1}^{2n}\left(D_{\tilde{x}_i}g + \frac{D_{\tilde{x}_i}^2 g}{2!} + \frac{D_{\tilde{x}_i}^3 g}{3!} + \frac{D_{\tilde{x}_i}^4 g}{4!} + ...\right) \quad (7)$$

Comparing this series with the true series, we see that different values for the predicted mean occur only if the moments of $\Delta x$ and $\tilde{x}_i$ are different. The distribution of $\tilde{x}_i$ is symmetric. All the odd moments are zero and hence all the odd terms sum to zero. Recalling that the sigma points are found from the column (or row) vectors of the matrix square root of $\sqrt{(n+k)P_{xx}}$, the second order even term is

$$\frac{D_{\tilde{x}_i}^2 g}{2!} = \left(\frac{\nabla^T \tilde{x}_i \tilde{x}_i^T \nabla}{2!}\right) \quad (8)$$

Therefore, the predicted mean is

$$\bar{y} = g[\bar{x}] + \frac{\nabla^T P_{xx} \nabla}{2!} + \frac{1}{2(n+k)} \sum_{i=1}^{2n} \left( \frac{D_{\bar{x}_i}^4 g}{4!} + ... \right) \quad (9)$$

Comparing equation (9) with true mean, shows that UT estimate mean due to 3$^{rd}$ order Taylor series and error accrue in forth and higher order terms. Equation (9) shows that the accuracy is affected by covariance and parameter k.
The true covariance of nonlinear function is given by

$$(P_{yy})_T = E\left[[y - \bar{y}_T][y - \bar{y}_T]^T\right] \quad (10)$$

Where the realization of the state error is

$$y - \bar{y}_T = g[\bar{x} + \Delta x] - \bar{y}_T$$
$$= D_{\Delta x} g + \frac{D_{\Delta x}^2 g}{2!} + \frac{D_{\Delta x}^3 g}{3!} + \frac{D_{\Delta x}^4 g}{4!} - E\left[\frac{D_{\Delta x}^2 g}{2!} + \frac{D_{\Delta x}^4 g}{4!} + ...\right] \quad (11)$$

Since $\Delta x$ is symmetric the expected value of the all odd order terms of $\Delta x$ evaluate to zero and the true covariance is

$$(P_{yy})_T = J_g P_{xx} J_g^T + E\left[\frac{D_{\Delta x} g (D_{\Delta x}^3 g)^T}{3!} + \frac{D_{\Delta x}^2 g (D_{\Delta x}^2 g)^T}{2 \times 2!} + \frac{D_{\Delta x}^3 g (D_{\Delta x} g)^T}{3!}\right]$$
$$- \left[\left(\frac{\nabla^T P_{xx} \nabla}{2!}\right) g\right]\left[\left(\frac{\nabla^T P_{xx} \nabla}{2!}\right) g\right]^T + ... \quad (12)$$

The linearization algorithm predicts the covariance using

$$(P_{yy})_{LIN} = J_g P_{xx} J_g^T \quad (13)$$

Although the first term was considered in respect to Taylor series, the later ones were neglected. However, the error accrues in fourth and higher order terms.
For UT the predicted covariance is

$$P_{yy} = J_g P_{xx} J_g^T + \frac{1}{2(n+k)} \sum_{i=1}^{2n} \left( \frac{D_{\bar{x}_i} g (D_{\bar{x}_i}^3 g)^T}{3!} + \frac{D_{\bar{x}_i}^2 g (D_{\Delta x}^2 g)^T}{2 \times 2!} + \frac{D_{\bar{x}_i}^3 g (D_{\bar{x}_i} g)^T}{3!} \right)$$
$$- \left[\left(\frac{\nabla^T P_{xx} \nabla}{2!}\right) g\right]\left[\left(\frac{\nabla^T P_{xx} \nabla}{2!}\right) g\right]^T + ... \quad (14)$$

Comparing (14) with true covariance shows that both linearization and UT are accurate to 3$^{rd}$ order terms but UT at least has same singed in forth and higher order term, so UT has less error estimation in covariance than EKF.
As it has been presented in (9) the mean of UT is dependent on covariance of state and since the accuracy of Linearization and UT prediction covariance has the same accuracy to 3$^{rd}$ order terms, the mean of UT doesn't change significantly in New UKF.

### III. SIGMMA POINTS PROPAGATION BASED ON UT AND LINEARIZED COVARIANCE

for priori mean $\bar{x}$ the sigma points are as below:

$$\chi_0 = \bar{x}$$
$$\chi_i = \bar{x} + (\sqrt{(n+\lambda)P_x})_i \quad i = 1,...,n \quad (15)$$
$$\chi_i = \bar{x} - (\sqrt{(n+\lambda)P_x})_{i-n} \quad i = n+1,...,2n$$

$\chi_i$ Transfer through nonlinear function h:

$$y_i = h(\chi_i) \quad i = 0,...,2n$$

And weighted mean for transfer sigma point is prediction step in Kalman filter.

$$\bar{y} \approx \sum_{i=0}^{2n} W_i^m y_i \quad (16)$$

For covariance we use (13) as linearized method that has the accuracy to 3$^{rd}$ order Taylor series. By this two steps for priori mean and covariance of sigma point the new estimation algorithm is:

---

*nonlinear function*:
$$x_k = f(x_{k-1}, u) + w_k$$
$$y_k = h(x_k) + v_k$$

*initialize with*:
$$\hat{x}_0 = E[x_0]$$
$$P_0 = E\left[(x_0 - \hat{x}_0)(x_0 - \hat{x}_0)^T\right]$$
$$\text{for } K \in \{1,...,\infty\}$$

*calculate sigma points*:
$$\chi_{k-1} = \left[\hat{x}_{k-1} \quad \hat{x}_{k-1} + \sqrt{(n+\lambda)P_{k-1}} \quad \hat{x}_{k-1} - \sqrt{(n+\lambda)P_{k-1}}\right]$$

*time update*:
$$\chi_{k|k-1} = f(\chi_{k-1}, u)$$
$$\hat{x}_k^- = \sum_{i=0}^{2n} W_i^m \chi_{k|k-1}$$
$$y_{k|k-1} = h(\chi_{k|k-1})$$
$$\hat{y}_k^- = \sum_{i=0}^{2n} W_i^m y_{k|k-1}$$

*measurement update*:
$$S_k = H_k P_k^- H_k^T + R_k$$
$$K_k = P_k^- H_k^T S_k^{-1}$$
$$\hat{x}_k = \hat{x}_k^- + K_k(y_k - \hat{y}_k^-)$$
$$P_k = (I - K_k H_k) P_k^-$$

---

Fig. 1. new estimation algorithm

#### A. Mean error in new algorithm

The purpose is to obtain mean of the transfer sigma point.

$$\bar{y}_u = \sum_{i=1}^{2n} W^i y^i \quad (17)$$

the Taylor series expansion of (17) should consider.

$$\bar{y}_u = \frac{1}{2n} \sum_{i=1}^{2n} (h(\bar{x}) + D_{\bar{x}} h + \frac{1}{2!} D_{\bar{x}}^2 h + \frac{1}{3!} D_{\bar{x}}^3 h + ...)$$
$$= h(\bar{x}) + \frac{1}{2n} \sum_{i=1}^{2n} (D_{\bar{x}} h + \frac{1}{2!} D_{\bar{x}}^2 h + ...) \quad (18)$$

Since The distribution of $\tilde{x}_i$ is symmetric the all odd moments are zero and hence all the odd terms sum to zero then it can be said that,

$$\sum_{j=1}^{2n} D_{\tilde{x}(j)}^{2k+1} h = \sum_{j=1}^{2n}\left[\left(\sum_{i=1}^{n} \tilde{x}_i^{(j)}\frac{\partial}{\partial x_i}\right)^{2k+1} h(x)\big|_{x=\tilde{x}}\right]$$

$$= \sum_{j=1}^{2n}\left[\left(\sum_{i=1}^{n} \tilde{x}_i^{(j)}\right)^{2k+1} \frac{\partial^{2k+1}}{\partial x_i^{2k+1}} h(x)\big|_{x=\tilde{x}}\right] \quad k>0 \quad (19)$$

$$= \sum_{j=1}^{n}\left[\left(\sum_{i=1}^{2n} \tilde{x}_i^{(j)}\right)^{2k+1} \frac{\partial^{2k+1}}{\partial x_i^{2k+1}} h(x)\big|_{x=\tilde{x}}\right]$$

$$= 0$$

Then …

$$\bar{y}_u = h(\bar{x}) + \frac{1}{2n}\sum_{i=1}^{2n}(\frac{1}{2!}D_{\tilde{x}}^2 h + \frac{1}{4!}D_{\tilde{x}}^4 h + \ldots)$$

$$= h(\bar{x}) + \frac{1}{2n}\sum_{i=1}^{2n}(\frac{1}{2!}D_{\tilde{x}}^2 h) + \frac{1}{2n}\sum_{i=1}^{2n}(\frac{1}{4!}D_{\tilde{x}}^4 h + \frac{1}{6!}D_{\tilde{x}}^6 h + \ldots) \quad (20)$$

The second Taylor series is extended as below

$$\frac{1}{2n}\sum_{i=1}^{2n}(\frac{1}{2!}D_{\tilde{x}}^2 h) = \frac{1}{2n}\sum_{i=1}^{2n}\frac{1}{2!}(\sum_{i=1}^{n} \tilde{x}_i^{(k)} \frac{\partial}{\partial x_i})^2 h(x)\big|_{x=\tilde{x}}$$

$$= \frac{1}{4n}\sum_{i=1}^{2n}\sum_{j=1}^{n} \tilde{x}_i^{(k)} \tilde{x}_j^{(k)} \frac{\partial^2}{\partial x_i \partial x_j} h(x)\big|_{x=\tilde{x}} \quad (21)$$

$$= \frac{1}{2n}\sum_{i,j=1}^{n}\sum_{k=1}^{n} \tilde{x}_i^{(k)} \tilde{x}_j^{(k)} \frac{\partial^2}{\partial x_i \partial x_j} h(x)\big|_{x=\tilde{x}}$$

Then by using the obtained sigma points, second Taylor series could be calculated

$$\frac{1}{2n}\sum_{i,j=1}^{n}\sum_{k=1}^{n} \tilde{x}_i^{(k)} \tilde{x}_j^{(k)} \frac{\partial^2 h(x)}{\partial x_i \partial x_j}\big|_{x=\tilde{x}} = \frac{1}{2n}\sum_{i,j=1}^{n}\sum_{k=1}^{n} (\sqrt{np})_{k_i}(\sqrt{np})_{k_j}\frac{\partial^2 h(x)}{\partial x_i \partial x_j}\big|_{x=\tilde{x}}$$

$$= \frac{1}{2n}\sum_{i,j=1}^{n}\sum_{k=1}^{n} nP_{i,j}\frac{\partial^2 h(x)}{\partial x_i \partial x_j}\big|_{x=\tilde{x}}$$

$$\bar{y}_u = h(\bar{x}) + \frac{1}{2}\sum_{i,j=1}^{n} P_{ij}\frac{\partial^2 h}{\partial x_i \partial x_j}\big|_{x=\tilde{x}} + \frac{1}{2n}\sum_{i=1}^{2n}(\frac{1}{4!}D_{\tilde{x}}^4 h + \frac{1}{6!}D_{\tilde{x}}^6 h + \ldots)$$

(22)

The above equation (22) is an extension of the mean of the proposed method. The error with true mean will be obtained using (23).

$$\bar{y} - \bar{y}_u = \left(E\left[\frac{1}{4!}D_{\tilde{x}}^4 h\right] + E\left[\frac{1}{6!}D_{\tilde{x}}^6 h\right] + \ldots\right) - \frac{1}{2n}\sum_{i=1}^{2n}(\frac{1}{4!}D_{\tilde{x}}^4 h + \frac{1}{6!}D_{\tilde{x}}^6 h + \ldots)$$

It is clear that in UT transform the mean error will produce in higher order Taylor series.

*B. Covariance error in new algorithm*

In (12), (13) the true and linearized covariance were shown then the difference is:

$$(P_{yy})_T - (P_{yy})_{LIN} = E\left[\frac{D_{\Delta x}g(D_{\Delta x}^3 g)^T}{3!} + \frac{D_{\Delta x}^2 g(D_{\Delta x}^2 g)^T}{2\times 2!} + \frac{D_{\Delta x}^3 g(D_{\Delta x} g)^T}{3!}\right] + \ldots \quad (23)$$

That shows it is accurate to 3rd order Taylor series like EKF and UKF. In this section, we show that the new algorithm is accurate to 3rd order Taylor series like UKF, however, UKF covariance at least has same singed term in fourth and higher order term, so UKF has less error estimation in covariance than EKF and New KF.

IV. COMPUTATION ANALYSIS OF NEW UKF

UKF has a higher computation cost in comparison to EKF because of the square root of the covariance matrix for generating 2n+1 sigma points, propagating sigma points, and predicting covariance for states, measurements, and cross covariance to update Kalman gain in each iteration.

Many articles [11], [12], show that Cholesky factorization has lower computation complexity $O(n^3/3)$ and is more stable to evaluate square root of the matrix comparing to other methods.

Next, we analyze the computational complexity in terms of the floating point operation for the proposed algorithm. The computation complexities of some basic equations are given in the table below according to [15] [2] [16].

$t_f$ Time required evaluating the function f

$t_{mm}$ Time required multiplying two $n\times n$ matrix

$t_m$ Time required multiplying a $n\times n$ matrix and $n\times 1$ matrix

$t_{ma}$ Time required multiplying a $n\times 1$ matrix and $1\times n$ matrix

$t_a$ Time required adding two $n\times 1$ matrixes

$t_{ms}$ Time required multiplying a scalar with a $n\times 1$ matrix

$t_{chol}$ Time required evaluating Cholesky factorization $n\times n$ matrix

$t_{tr}$ Time required for transpose matrix

$t_{am}$ Time required adding two $n\times n$ matrixes

If consider $t_{sa}, t_{ms}$ as basic operation, we can write:

$$t_{max} \geq max\{t_{sa}, t_{ms}, t_{sd}\}$$
$$t_{mm} \leq n^3 t_{max}$$
$$t_m \leq n^2 t_{max}$$
$$t_{ma} \leq n t_{max}$$
$$t_{chol} \leq n^3/3 t_{max} \quad (24)$$
$$t_a \leq n t_{max}$$
$$t_{md} \leq (n+1) t_{max}$$
$$t_{tr} \leq n^2 t_{max}$$

If number of basic operation required evaluating the function $f$ is $j(j \in \mathbb{N})$ then,

$$t_f \leq j t_{max} \quad (25)$$

So the computation complexity of UKF for 2n+1 sigma point should be

$$t_{ukf} = (j + 8m + 18n + 2jn + 10mn + 8m^2 n$$
$$+ 10m^2 + 4m^3 + 28n^2 + 12n^3 + 4)t_{max} \quad (26)$$
$$t_{pukf} = (j + 5n + 2jn + mn + m^2 + 13n^2 + 5n^3)t_{max} \quad (27)$$

Numerical analysis shows that SPUKF algorithm can reduce the computation cost about 50% for systems with state number less than 10. But it can be different according to $n$, $m$ and $J$ of the nonlinear system. Figure 3 shows that for a big scale system this algorithm reduces the computation cost more than 50%. For systems with $m$ near $n/2$ and $J=10*n$ it can reduce the computation cost about 75%. Therefore, unlike reduced sigma point UKFs algorithm and SPUKF the new KF is efficient even for big scale system.

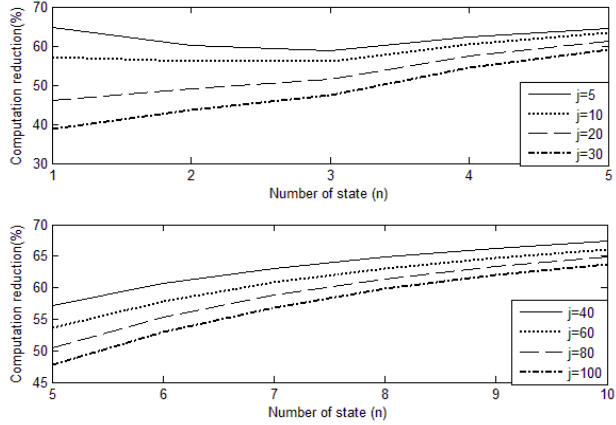

Fig. 2. computation reduction in UKF with different n, J

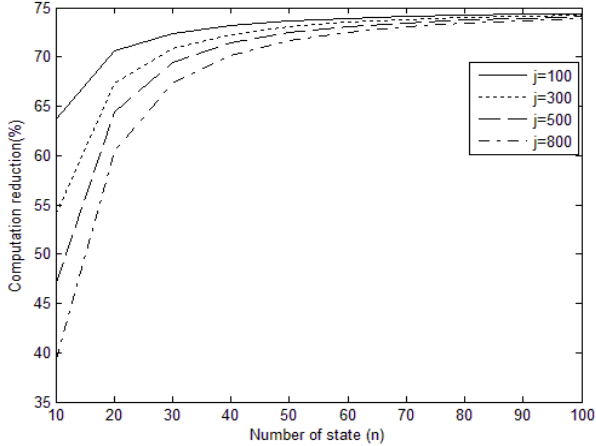

Fig. 3. computation reduction in UKF with different n, J for big scale systems

## V. EXAMPLE

In this section we demonstrate the performance of HUKF and compare it to UKF and EKF.

We choose two systems as benchmark.
A- A time series with nonlinearity in both process model and measurement, which has been used in my article.

$$x_{t+1} = 1 + \sin(\omega \pi t) + \phi x_t + v_t \quad (28)$$

Where $v_t$ is Gamma(2,3) random variable for modeling process noise and "$\omega = 4e-2$, $\phi = 0.5$" are scaler parameters. A nonstationary observation model,

$$y_t = \begin{cases} \phi x_t^2 + n_t & t \leq 30 \\ \phi x_t - 2 + n_t & t > 30 \end{cases} \quad (29)$$

Is used. The observation noise, $n_t$ is zero mean Gaussian distribution. The simulation repeated 1000 times. Table 1 summarize the performance and execution time of Nonlinear filters. The state and the error covariance for the filter initialization are:

$x(0) = 1;$  $x_h(0) = 1;$
$P = I * 10^{-3}$  particle number=200

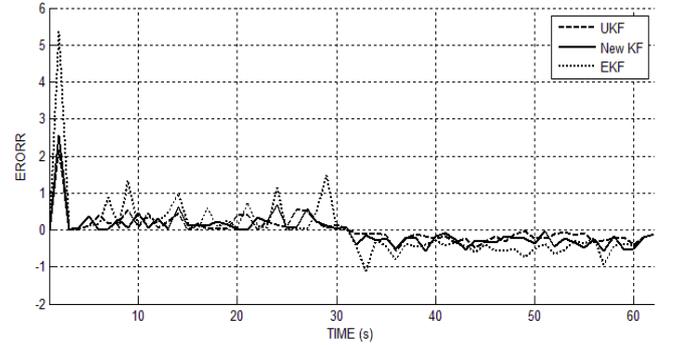

Fig. 4. x1 error of EKF, UKF and HUKF

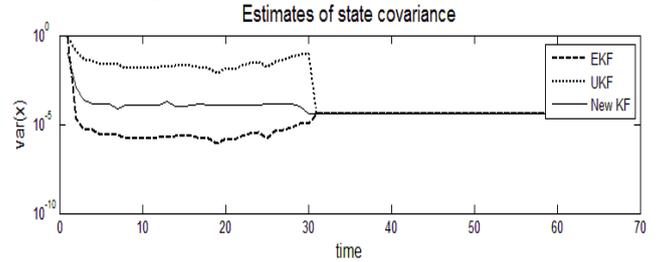

Fig.5. State covariance

For this example, the computation reduction of new algorithm according to (26) (27) is

$$\frac{t_{ukf} - t_{nkf}}{t_{ukf}} = 65\% \quad \text{for n=1, m=1, j=5}$$

In table below the Time requirement and MSE of estimation of different filter method were investigated for 1000 simulation.

TABLE 1: filters mean MSE and execution time for 1000 independent simulations.

| FILTER | Execution Time (s) | MSE (mean) |
|---|---|---|
| EKF | 0.0641 | 0.387 |
| SSUKF | 0.1428 | 0.302 |
| SPUKF | 0.1827 | 0.316 |
| UKF | 0.2332 | 0.271 |
| **New KF** | **0.1194** | **0.279** |
| PF | 0.7114 | 0.437 |

| PF-EKF | 1.4861 | 0.293 |
|---|---|---|
| PF-UKF | 3.8732 | 0.090 |
| **PF-New KF** | **1.7807** | **0.113** |

The table shows that New UKF almost has the same accuracy to UKF for this example, and reduce Execution time more than SSUKF and SPUKF, but without increasing estimation error, since the function f is not a complicated one (j=6), the SPUKF is not efficient enough. In addition to that, it is accurate to first order Taylor series. SSUKF and HUKF decrease time requirement remarkably, however, New KF has better accuracy and it is more suitable to implement in real application due to its simplicity. since process noise is non Gaussian we use particle filter with New KF as proposal distribution function. results show that New KF reduces time consumption in comparison with EPF[2] and UPF[3]. the reason behind the better operation of UKF and New KF methods is the usage of nonlinear function of the system for transformation the sigma points. Also EKF error in the first period (0-30s), is due to the nonlinear observation model of time series which is obviously, far more than the estimation error in second period (30-60 s).

B- State and parameter estimation in hybrid maglev:

One of the reasons for developing new nonlinear Kalman filter was to implement sensorless methods to control Amirkabir industrial laboratory" hybrid maglev" (AIL hybrid maglev). This novel hybrid permanent magnet-electro magnet suspension system has an optimal structural design to decrease power loss about 90% [13]. The first priority of State estimation for hybrid maglev is an estimation of vertical velocity which is difficult to measure and the second is to estimate other states to increase the reliability of this type of transportation system in the presence of sensor malfunction. Moreover, since the weight of this system (Load) is variable, parameter estimation is important to reduce the error of estimation and regulation of air gap. Consider nonlinear augmented state equation of hybrid maglev (30) for state and parameter estimation.

$$\dot{x}_1(t) = x_2(t)$$
$$\dot{x}_2(t) = -\frac{\mu_0 A_{ag}(Nx_3 - H_C L_{PM})^2}{m(2x_1(t) + \frac{L_{PM} A_{ag}}{\mu_r A_{PM}} + R_C(\frac{h}{R_L} + \mu_0 A_{ag}) + \frac{hL_{PM}}{\mu_0 \mu_r A_{PM} R_L})^2} + g \quad (30)$$
$$\dot{x}_3(t) = \frac{x_2(t)x_3(t)}{x_1(t)} - \frac{Rx_1(t)x_3(t)}{K} + \frac{x_1(t)u(t)}{K}$$
$$\dot{m} = w$$

Here m is weight of maglev and w is artificial disturbance use in augmented Kalman filters. For this example, the computation reduction of new algorithm according to (26) (27) is

$$\frac{t_{ukf} - t_{nkf}}{t_{ukf}} = 61\% \quad \text{for n=4, m=1, j=30}$$

---

[2] Extended particle filter

[3] Unscented particle filter

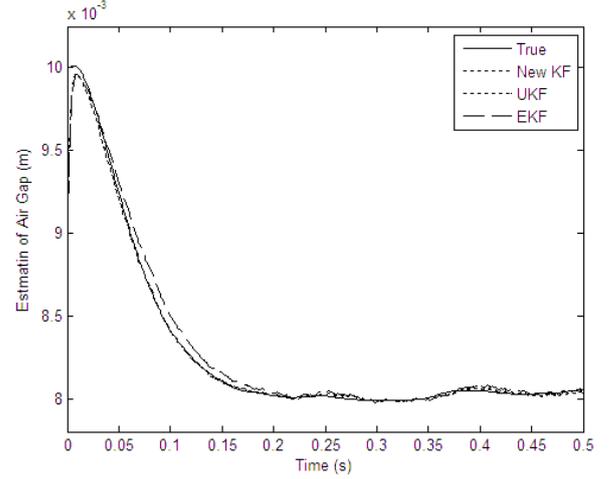

*Fig. 6. parameter estimation in hybrid maglev train*

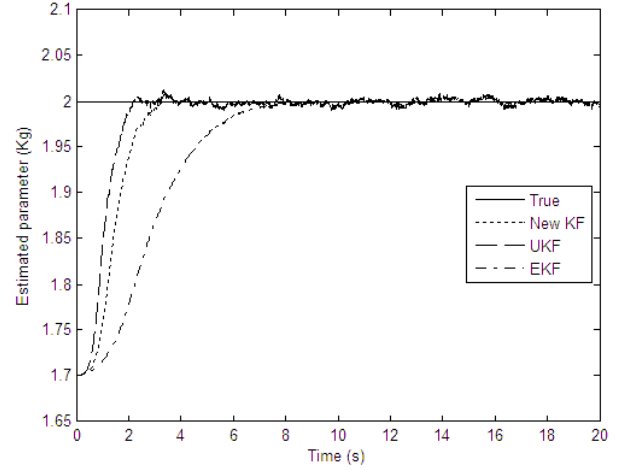

*Fig. 7. Parameter estimation of hybrid maglev train*

*Table 2: Estimations MSE for different KFs*

| Filter | Parameter estimation MSE | Air gap estimation MSE |
|---|---|---|
| UKF | 0.1334 | 1.867e-6 |
| New KF | 0.1568 | 1.935e-6 |
| EKF | 0.6471 | 2.346e-6 |

For Air gap estate estimation EKF, UKF and New KF almost have same accuracy except in transient time, however in parameter estimation UKF and New KF significantly provides better result than EKF because of convergence speed and their accuracy to 3$^{rd}$ order Taylor series. It is be noted that the execution time of New KF is about $0.5\,T_{ukf}$, so in this situation which both have same performance the New KF is better choice.

## VI. CONCLUSION

The New KF can reduce computation cost about 50% and 75% for big scale system without stability problem, and lack of accuracy. It also can estimate mean and covariance to 3rd order Taylor series comparing to true mean and covariance of the nonlinear function. Simulations and numerical analysis show that New KF has superior performance to EKF and simple to implement for both discrete and continuous system even with nonlinearity in measurements. In addition to above mentioned advantages, New KF neither has the problems that the reduce sigma points have, nor decreases the accuracy of filter like SPUKF. Table (3) summarizes the characteristics of different methods.

Table 3: comparison between different methods that proposed for computation reduction of UKF.

| Filter | TIME | Mean accuracy (Taylor series) | Covariance accuracy (Taylor series) | Numerical stability |
|---|---|---|---|---|
| **New KF** | $0.3$-$0.5\, T_{ukf}$ | 3rd order | 3rd order | Stable |
| **SPUKF** | $0.5$-$0.8\, T_{ukf}$ | 1st order | 1st order | Stable |
| **RSUKFs** | $0.5\, T_{ukf}$ | 2st order | 2st order | Unstable |
| **MUT** | variable | 2st order | 2st order | stable |

From the facts that mentioned in this paper we can say that this new Kalman filter is a better choice in real time applications than UKF and EKF and other methods that have been presented to reduce the computation cost of UKF due to its accuracy and simplicity.